\documentclass[aps,pra,twocolumn,superscriptaddress,showpacs,showkeys,amsmath,amssymb]{revtex4}

\usepackage{amsfonts}
\usepackage{amssymb,amsmath}
\usepackage{mathrsfs}
\usepackage{latexsym}
\usepackage{amsmath}
\usepackage[cp1251]{inputenc}
\usepackage{graphicx}
\usepackage{dcolumn}
\usepackage{bm}
\usepackage{color}

\RequirePackage{ifthen}
\RequirePackage[pdfstartview=FitH]{hyperref}
\begin{document}
	
	\title{Competition of superfluid phases in low-dimensional spin-$1\over 2$ fermions\\ with $s$- and $p$-wave interactions}
	\author{A.~Nikolaeva} \affiliation{Professor Ivan Vakarchuk Department for Theoretical Physics, Ivan Franko National University of Lviv, 12 Drahomanov Street, Lviv, Ukraine}
	\author{O.~Hryhorchak} \affiliation{Professor Ivan Vakarchuk Department for Theoretical Physics, Ivan Franko National University of Lviv, 12 Drahomanov Street, Lviv, Ukraine}
	\author{V.~Pastukhov\footnote{e-mail: volodyapastukhov@gmail.com}}
	\affiliation{Professor Ivan Vakarchuk Department for Theoretical Physics, Ivan Franko National University of Lviv, 12 Drahomanov Street, Lviv, Ukraine}

	\date{\today}

	\pacs{67.85.-d}
	
	\keywords{1D superfluid fermions, $s$-wave and $p$-wave contact potentials, BCS-BEC crossover}
	
	\begin{abstract}
	The ground state of spin-$1\over 2$ fermions with contact $s$-wave inter- and $p$-wave intra-species interactions is discussed. Particularly, we formulate the mean field scheme for calculating thermodynamic properties of the system in arbitrary dimension $D<2$ and discuss in detail the phase diagram in 1D case. Except clean phases with either singlet or triplet Cooper pairings, we have identified two mixed phases (one stable and another metastable) of the one-dimensional two-component fermions where both pairing mechanisms coexist.
	\end{abstract}
	
\maketitle

\section{Introduction}

The experimental realization of ultra-cold Bose \cite{Anderson_1995,Bradley_1995,Davis_1995} and Fermi \cite{DeMarco_1999,Truscott_2001,Schreck_2001} gases more than 25 years ago stimulated further development in the field. The most promising from the perspective of the phase-diagram richness are one-dimensional systems of a few \cite{Sowiski_2019} and macroscopic number \cite{Imambekov_2012} of particles. An important feature of the ultra-cold atomic gases is their universality, when low density provides insensibility of properties of these systems to any details of microscopic two-body potential. The latter, therefore, can be replaced by a local $\delta$-function-like pseudopotential. Furthermore, below 2D the finite-range corrections to the two-body point-like interaction are less relevant (in the renormalization group sense) than the contact three-body interaction. This fact predetermines \cite{Pastukhov_2019,Valiente_2019,Morera_2022} an universal many-body properties of bosons with resonant two-body interaction in 1D and the Efimov-like effect in the four- and five-body sectors in higher fractional dimensions \cite{Hryhorchak_2022}. The situation is less diverse in a case of fermions: the contact potential is only possible for particles in different spin states, while the minimal two-body interaction between identical fermions is of $p$-wave type; the Pauli principle permits the three-body interaction only for a systems with higher symmetries starting from $SU(3)$ \cite{Drut_2018}.

In last few years, physics of identical fermions in 1D with $p$-wave interaction ($\delta''$ pseudopotential) attracted much attention in the literature \cite{Tajima_2021,Tajima_2022,Tanaka_2022,Maki_2023}. Being renormalizable in the two-body sector in 1D \cite{Cui_2016} and fractional \cite{Pastukhov_2020} dimensions below $D=2$, this local interaction effectively models the finite-range two-body potential in a case of identical fermions. The phase diagram of identical fermions (see, \cite{Gurarie} and \cite{Yang_2020} for 3D and 2D cases, respectively) in $D>1$ intrinsically contains \cite{Pastukhov_2020} the crossover region between the Bardeen-Cooper-Schrieffer (BCS) pairing mechanism and the Bose-Einstein condensation (BEC) of dimers. The universal character of the effective $p$-wave interaction allows for existence \cite{Cui_Dong_2016,Sekino_2018} of the exact identities that relate the thermodynamic functions of many-body system with the high-momentum tail of particle distribution. The normal state properties of 1D fermions with $p$-wave interaction were studied in Ref.~\cite{Maki_2021} using the high-temperature series expansion. A detailed variational treatment of two- and three-body states in the system of macroscopic number of identical Fermi particles is presented in Ref.~\cite{Guo_2022}.
No less interesting aspect of the one-dimensional world is the existence of an exact duality between behavior of spin-polarized fermions with the $p$-symmetric interaction and a system of bosons with $\delta$-repulsion \cite{Valiente_2020,Valiente_2021,Sekino_2021}. This amazing fact allows \cite{Granet_2022} the perturbative analysis of strongly-interacting bosons in terms of standard fermionic diagrammatic technique. 

The objective of this paper is to reveal the competing of pairing mechanisms for the low-dimensional ($D<1$) spin-$1\over 2$ fermions with the $s$-wave and $p$-wave interactions between two particles in a different and in a same spin states, respectively. Recently, properties of a somewhat similar model, without $p$-wave interaction between particles of one sort, were discussed \cite{Guo_2023} in context of the pairing-tripling competing.

\section{Model description}
We consider the system of $N$ spin-$1\over 2$ (with projections on the quantization axis denoted by $\sigma=\uparrow,\downarrow$) interacting fermions loaded in the $D$-dimensional box of size $L^D$ with periodic boundary conditions. The interaction between two spin-up--spin-down ($\uparrow\downarrow$) particles is assumed to be of zero range and attractive with coupling constant $g_{s,\Lambda}$. The model also takes into account the minimal local (in real space) $p$-wave type interaction $(\nabla \delta({\bf r}))\nabla$ of $\uparrow\uparrow$ and $\downarrow\downarrow$ fermions \cite{Pastukhov_2020} which is characterized by the cutoff-dependent bare coupling $g_{p,\Lambda}$. The grand-canonical Hamiltonian, which commutes with both $N_{\uparrow}$ and $N_{\downarrow}$, is specified as follows
\begin{eqnarray}\label{H}
&& H-\mu N=\sum_{{\bf p}, \sigma}\xi_{p}\psi^{\dagger}_{\sigma,{\bf p}}\psi_{\sigma,{\bf p}}\nonumber\\
&&-\frac{g_{s,\Lambda}}{L^D}\sum_{{\bf p},{\bf k},{\bf q}}\psi^{\dagger}_{\uparrow,{\bf p}+{\bf k}}\psi^{\dagger}_{\downarrow,{\bf p}-{\bf k}}\psi_{\downarrow,{\bf p}-{\bf q}}\psi_{\uparrow,{\bf p}+{\bf q}}\nonumber\\
&&+\frac{g_{p,\Lambda}}{2L^D}\sum_{{\bf p},{\bf k},{\bf q}, \sigma}{\bf k}{\bf q}\psi^{\dagger}_{\sigma,{\bf p}+{\bf k}}\psi^{\dagger}_{\sigma,{\bf p}-{\bf k}}\psi_{\sigma,{\bf p}-{\bf q}}\psi_{\sigma,{\bf p}+{\bf q}},
\end{eqnarray}
where $\psi^{\dagger}_{\sigma,{\bf p}}$ ($\psi_{\sigma,{\bf p}}$) are the fermionic creation (annihilation) operators; $\xi_{p}=\varepsilon_p-\mu=\frac{\hbar^2 p^2}{2m}-\mu$, and all summations over the wave-vectors are restricted by the ultraviolet (UV) cutoff $\Lambda$. In order to find out the correct renormalization of the $s$-wave and $p$-wave coupling constants, it is necessary to consider the two-body problem in vacuum.

\subsection{Overview of the two-body problem}
The two-fermion states of Hamiltonian $H$ can be conventionally classified by the value of total spin $S$. There is exactly one bound state in the spin-zero (singlet) sector and the two-fold generated energy levels (with $S_z=\pm 1$) in the spin-one (triplet) sector. 

An eigenvalue of Hamiltonian $H$, $\epsilon_s=-\frac{\hbar^2}{ma^2_s}$, corresponding to the spherically-symmetric part of a singlet bound state $|s\rangle=\sum_{{\bf k}}A_s(k)\psi^{\dagger}_{\uparrow,{\bf k}}\psi^{\dagger}_{\downarrow,-{\bf k}}|\textrm{vac}\rangle$ satisfies the following equation
\begin{eqnarray}
g^{-1}_{s,\Lambda}=\frac{1}{L^D}\sum_{{\bf k}}\frac{1}{2\varepsilon_k+|\epsilon_s|}.
\end{eqnarray}
When $D<2$, the sum in r.h.s. is convergent and consequently a coupling constant $g_{s,\Lambda}$ needs no regularization (i.e. is independent of $\Lambda$)
\begin{eqnarray}
g^{-1}_{s,\Lambda}=g^{-1}_{s}=\frac{\Gamma(1-D/2)}{(4\pi)^{D/2}}\frac{ma^{2-D}_s}{\hbar^2},
\end{eqnarray}
while the appropriate wave function remains universal up to $D<4$ and is given by the Lorentzian $A_s(k)\propto \frac{1}{1+k^2a^2_s}$ in momentum space.

The spin-one sector, where the spinor part is symmetric, is characterized by the odd-wave spacial part of the wave function. The simplest non-trivial, in context of our model (\ref{H}), form is the $p$-wave symmetric $|p\rangle=\sum_{{\bf k}}A_p({\bf k})\psi^{\dagger}_{\sigma,{\bf k}}\psi^{\dagger}_{\sigma,-{\bf k}}|\textrm{vac}\rangle$ [where $A_p({-\bf k})=-A_p({\bf k})$]. By the successive substitution in the Schr\"odinger equation one obtains the $p$-symmetric wave function $A_p({\bf k})\propto \frac{{\bf k}{\bf n}}{1+k^2a^2_p}$ (here ${\bf n}$ is a unit constant vector) with the bound-state energy $\epsilon_p=-\frac{\hbar^2}{ma^2_p}$. Note that $A_p({\bf k})$ is square-integrable only when $D<2$. For higher dimensions the universality of the $p$-wave bound-state wave function is broken and the two-body properties of the system depend on the details of potential at short distances. The eigenvalue problem reduces to the solution of equation
\begin{eqnarray}
g^{-1}_{p,\Lambda}+\frac{1}{L^D}\sum_{{\bf k}}\frac{k^2/D}{2\varepsilon_k+|\epsilon_p|}=0.
\end{eqnarray}
A sum over the wave vector in the above equation crucially depends on the UV cutoff.
Introducing renormalized (observable) coupling constant $g^{-1}_{p,\Lambda}+\frac{1}{L^D}\sum_{{\bf k}}\frac{m}{D\hbar^2}=g^{-1}_{p}$, one can relate $g_p$ (if $g_p>0$) to the two-body bound state energy in $D<2$
\begin{eqnarray}
g^{-1}_{p}=\frac{\Gamma(1-D/2)}{(4\pi)^{D/2}D}\frac{m}{\hbar^2a^D_p}.
\end{eqnarray}
Below, the case of negative coupling constant $g_{p}$ will be also parametrized by $a_p$. The impact of the $p$-wave interaction on the energy levels of low-momentum scattering states of two identical fermions is the following: $g_{p}/L^D$. Therefore, a positive $g_{p}$ increases the energy of the system and can be identified as `repulsive' interaction, while the case of negative $g_{p}$s should be thought as the `attractive' one. This situation is somewhat similar to the two-body problem with $s$-wave pseudopotential in $2\le D<4$ \cite{Hryhorchak_2023}.

The $p$-wave interaction presented in model (\ref{H}) is too singular for the adequate description of the three-body problem. Even in 1D, the effective-field-theory Hamiltonian $H$ should be supplemented \cite{Sekino_2021} by the higher-order interaction terms. The solution of this issue in $D>1$ will be given elsewhere since details of the three-body physics of the adopted model do not affect our further consideration.

\section{Mean-field consideration}
Let us briefly discuss possible phases of the considered system in the many-body limit. The attractive character of the $s$-wave interaction between two fermions that are in different spin states, suggests the Cooper paring with the formation of the singlet-pairs BEC at zero temperature in dimensions $D>1$. On the other hand, the presence of the $p$-wave interaction between identical Fermi particles can lead to the formation of two-body pairs in this channel. Depending on a sign of coupling $g_p$, the nature of pairing mechanisms is quite different. For a negative values of the renormalized coupling $g_p$, the $p$-wave channel BCS mechanism realizes, while for the `repulsive' case $g_p>0$ the two-body bound states with the vacuum origin occur. In the system of $D$-dimensional spin-polarized fermions these two regimes meet each other when $g^{-1}_p$ is strictly positive, signaling the BCS-BEC crossover \cite{Pastukhov_2020} behavior. 

From which it follows, we should expect the coexistence of various superfluid phases in the system described by Hamiltonian $H$. While simplifying (\ref{H}) in the spirit of the mean field (MF) approximation and in order to display all pairing mechanisms occurring in our system correctly, we transform our initial model as follows ($H\to H_{MF}$):
\begin{eqnarray}\label{}
&&H_{MF}-\mu N=\sum_{{\bf k}, \sigma}\xi_{k}\psi^{\dagger}_{\sigma,{\bf k}}\psi_{\sigma,{\bf k}}\nonumber\\
&&-\frac{g_{s}}{L^D}\sum_{{\bf k},{\bf q}}\left[\psi^{\dagger}_{\uparrow,{\bf k}}\psi^{\dagger}_{\downarrow,-{\bf k}}\Delta_{{\bf q}}+\textrm{h.c.}\right]\nonumber\\
&&+\frac{g_{p,\Lambda}}{2L^D}\sum_{{\bf k},{\bf q}, \sigma}{\bf k}{\bf q}\left[\psi^{\dagger}_{\sigma,{\bf k}}\psi^{\dagger}_{\sigma,-{\bf k}}\eta_{\sigma,{\bf q}}+\textrm{h.c.}\right]\nonumber\\
&&+\frac{g_{s}}{L^D}\sum_{{\bf k},{\bf q}}\Delta^*_{{\bf k}}\Delta_{{\bf q}}-\frac{g_{p,\Lambda}}{2L^D}\sum_{{\bf k},{\bf q}, \sigma}{\bf k}{\bf q}\eta^*_{\sigma,{\bf k}}\eta_{\sigma,{\bf q}},
\end{eqnarray}
where last two constant terms provide an appropriate expectation value of $H_{MF}$. The anomalous averages 
\begin{eqnarray}\label{anomal_av}
\Delta_{{\bf q}}=\langle \psi_{\downarrow,-{\bf q}}\psi_{\uparrow,{\bf q}}\rangle, \ \ \eta_{\sigma,{\bf q}}=\langle \psi_{\sigma,-{\bf q}}\psi_{\sigma,{\bf q}}\rangle,
\end{eqnarray}
appeared because of the spontaneously broken initial global $U(1)\times U(1)$ symmetry and should be calculated self-consistently on the eigenstates of $H_{MF}$. Making use of notation for the (in general complex) scalar $\Delta$ and vector ${\bf \eta}_{\sigma}$ order parameters
\begin{eqnarray}\label{gap_Eq}
\Delta=\frac{g_s}{L^D}\sum_{{\bf q}}\Delta_{{\bf q}}, \ \ {\bf \eta}_{\sigma}=\frac{g_{p,\Lambda}}{L^D}\sum_{{\bf q}}{\bf q}\eta_{\sigma,{\bf q}},
\end{eqnarray}
one can rewrite the MF Hamiltonian by using compact matrix notations $\hat{\psi}_{\bf k}=(\psi_{\uparrow,{\bf k}}, \psi^{\dagger}_{\uparrow,-{\bf k}}, \psi_{\downarrow,{\bf k}}, \psi^{\dagger}_{\downarrow,-{\bf k}})^{T}$ as follows
\begin{eqnarray}\label{H_MF}
H_{MF}-\mu N=\frac{1}{2}\sum_{{\bf k}}\hat{\psi}^{\dagger}_{\bf k}\hat{h}_{\bf k}\hat{\psi}_{\bf k}+\sum_{{\bf k}}\xi_k\nonumber\\
+L^D\frac{|\Delta|^2}{g_{s}}-L^D\sum_{\sigma}\frac{|{\bf \eta}_{\sigma}|^2}{2g_{p,\Lambda}},
\end{eqnarray}
where the hermitian $4\times 4$ matrix $\hat{h}_{\bf k}$, reads
\begin{eqnarray}\label{}
\hat{h}_{\bf k}=\left(\begin{array}{c c c c}
\xi_k & {\bf k}{\bf \eta}_{\uparrow} & 0 &-\Delta\\
{\bf k}{\bf \eta}^*_{\uparrow} &-\xi_k & \Delta^* & 0\\
0 & \Delta & \xi_k & {\bf k}{\bf \eta}_{\downarrow}\\
-\Delta^* & 0 & {\bf k}{\bf \eta}^*_{\downarrow} &-\xi_k\\
\end{array}\right).
\end{eqnarray}
A non-zero value of $\Delta$ or $\eta_{\sigma}$ signals the emergence of superfluidity in the system. Particularly, a quantity $\Delta\neq 0$ in dimensions $D>1$ is responsible for BEC of the $s$-wave $\uparrow \downarrow$ dimers (singlets) at absolute zero of temperature, while $\eta_{\sigma,{\bf q}}\neq 0$ indicates the Bose condensation of the $p$-wave ($\uparrow \uparrow$ and $\downarrow \downarrow$) pairs (triplets). Exactly in 1D there are no BECs, but the system supports superfluidity of the Berezinskii-Kosterlitz-Thouless type with the off-diagonal quasi-long-range order of the dimer-dimer (both singlets and triplets) equal-time propagator. The MF approximation treats $s$- and $p$-wave pairs as a non-interacting Bose gasses, therefore, magnitudes of order parameters (\ref{gap_Eq}) are non-zero even in 1D. In principle, the richest phase that can be observed in our model is the coexistence of three superfluid components at most.

The Hamiltonian (\ref{H_MF}) can be diagonalized by means of unitary transformation. New annihilation operators $d_{\bf k}$ and $u_{\bf k}$ of two types of quasiparticles determine the MF ground state $u_{\bf k}|0\rangle=d_{\bf k}|0\rangle=0$. In terms of these operators the Hamiltonian (\ref{H_MF}) takes the form
\begin{eqnarray}\label{}
H_{MF}-\mu N=\sum_{{\bf k}}\sum_{\alpha=d,u}\mathcal{E}_{\alpha,{\bf k}}\alpha^{\dagger}_{\bf k}\alpha_{\bf k}+\Omega_{MF},
\end{eqnarray}
where we have introduced notation for two branches of excitation spectrum [assuming that $\Delta$ and ${\bf \eta}_{\sigma}$ are real and introducing ${\bf \eta}_{\pm}=\frac{1}{2}\left({\bf \eta}_{\uparrow}\pm {\bf \eta}_{\downarrow}\right)$]
\begin{eqnarray}\label{E_k}
\left.\begin{array}{c}
\mathcal{E}^2_{u,{\bf k}}\\
\mathcal{E}^2_{d,{\bf k}}\\
\end{array}
\right\}=\xi^2_k+\left[\sqrt{\Delta^2+({\bf k}{\bf \eta}_{+})^2}\pm |{\bf k}{\bf \eta}_{-}|\right]^2,
\end{eqnarray}
and for the ground-state contribution to the grand potential
\begin{eqnarray}\label{Omega_MF}
\Omega_{MF}=\sum_{{\bf k}}\left[\xi_k-\frac{1}{2}\sum_{\alpha=d,u}\mathcal{E}_{\alpha,{\bf k}}\right]\nonumber\\
+L^D\frac{|\Delta|^2}{g_{s}}-L^D\sum_{\sigma}\frac{|{\bf \eta}_{\sigma}|^2}{2g_{p,\Lambda}}.
\end{eqnarray}
Now we are able to calculate the anomalous averages at zero temperature $\langle\dots\rangle=\langle 0|\dots|0\rangle$ in Eqs.~(\ref{anomal_av}) to determine order parameters (\ref{gap_Eq}). There is, however, an equivalent and much simpler way 
for obtaining the gap equations by means of grand-potential minimization
\begin{eqnarray}\label{Omega_min}
\frac{\partial \Omega_{MF}}{\partial \Delta}=0, \ \ \frac{\partial \Omega_{MF}}{\partial {\bf \eta}_{\sigma}}=0.
\end{eqnarray}
These equations have to be supplemented by the thermodynamic identity $-\frac{\partial \Omega_{MF}}{\partial \mu}=N$ that relates chemical potential of the system to the total number of fermions.

\section{Phase diagram}
In general, the arbitrary-$D$ phase diagram is very rich and complicated, since the complex-valued vector order parameters ${\bf \eta}_{\sigma}$ contain $2D$ free parameters each. This general case deserves separate publication, and below we mainly focus on properties of the 1D system. Furthermore, taking into account the symmetry arguments it is natural to suggest that amplitudes of complex numbers ${\eta}_{\uparrow,\downarrow}$ are equal to each other in 1D, and the only difference is in the phase factor (let's say ${\eta}_{\downarrow}={\eta}_{\uparrow}e^{i\theta}$, while choosing, without loss of generality, both ${\eta}_{\uparrow}$ and $\Delta$ real and positive definite). Plugging the ansatz into $\Omega_{MF}$ and making use of the minimization procedure one is left (except trivial $\Delta=0$ or ${\eta}_{\uparrow,\downarrow}=0$ phases) with two possible values of the phase factor $e^{i\theta}=\pm 1$. From the point of view of excitation spectrum (\ref{E_k}), the `plus' sign ($\eta_{-}=0$ in this case) leads to the symmetric phase with twofold degenerated branches $\mathcal{E}_{u,{\bf k}}=\mathcal{E}_{d,{\bf k}}$ of quasiparticle excitations, while the `minus' sign of the phase factor refers to the antisymmetric phase, where ${\eta}_{\downarrow}=-{\eta}_{\uparrow}$ (and consequently $\eta_{+}=0$). Note that both phases are characterized by the simultaneous presence of singlet and triplet Cooper pairs, however, their thermodynamic properties are quite different. From the energy arguments it is also understood that large discrepancy between the two-body binding energies in the $s$- and $p$-channel makes the phase with only the singlet or triplet pairing more robust. It particularly means that the singlet phase is realized in the $a_s\ll a_p$ limit, while the $p$-wave pairs are more preferable in the opposite limit $a_s\gg a_p$. Of course, the most interesting region is $a_s \sim a_p$, where all four phases of the system can interchange each other. Because of zero temperature, the Gibbs phase rule for our two-component system allows only three of them to coexist (and consequently at most triple points are available on the phase diagram).

The numerical calculations of the phase diagram were performed with the assumption that the system is under a constant pressure rather than being loaded in a fixed volume. These are the most natural conditions from the point of view of experimental realization. Additionally, constant external pressure conditions -- where the equilibrium is characterized by minimum of the Gibbs free energy -- simplifies the numerical analysis, because throughout the possible metastable phases of the considered system the one with minimal chemical potential (Gibbs free energy per particle) is thermodynamically stable.

\begin{figure}
\centerline{\includegraphics
[width=0.55\textwidth,clip,angle=-0]{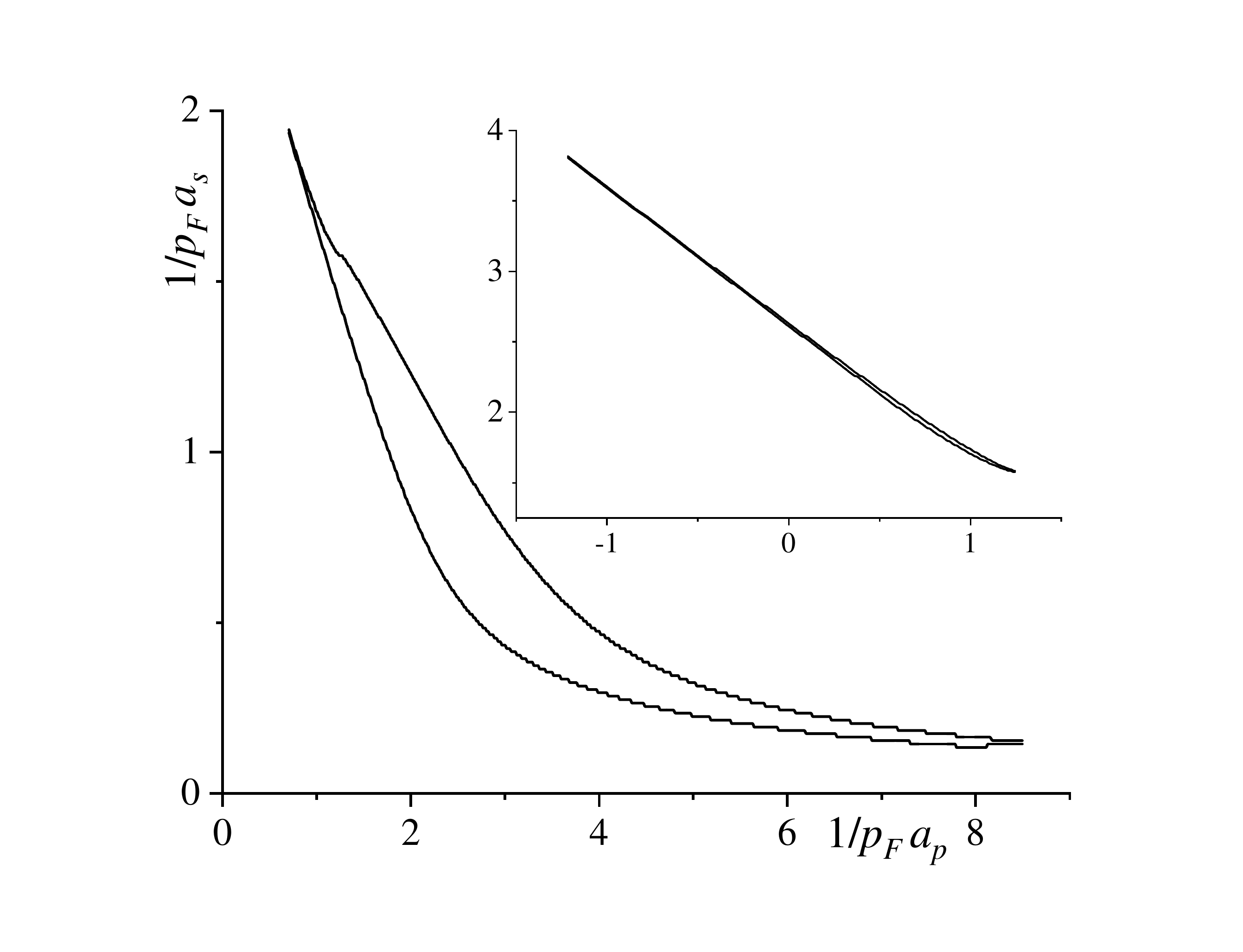}}
\caption{The one-dimensional MF region of existence of metastable antisymmetric phase, where the triplet order parameters being out of phase ${\eta}_{\downarrow}=-{\eta}_{\uparrow}$. Main panel and insert delimit areas, where the nontrivial solutions to coupled Eqs.~\ref{Omega_min} can be found from the BEC and BCS sides of the BCS-BEC crossover, respectively.}\label{anti-symm_phase_fig}
\end{figure}

\begin{figure}
\centerline{\includegraphics
[width=0.55\textwidth,clip,angle=-0]{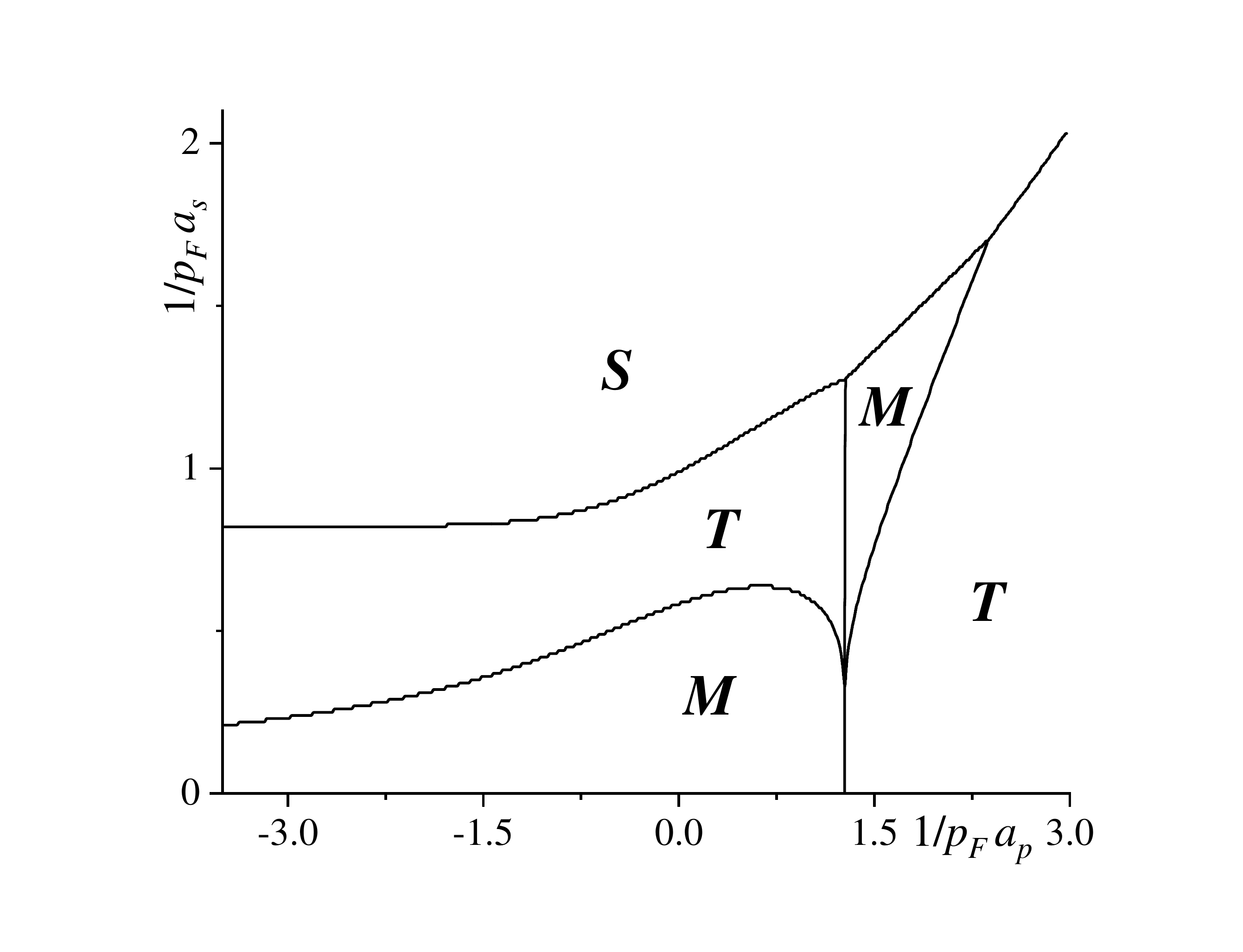}}
\caption{The MF phase diagram of the model described by Hamiltonian (\ref{H}) in 1D. $S$ denotes region with only singlet $\uparrow\downarrow$ pairing, $T$ -- region with the triplet $\uparrow\uparrow+\downarrow\downarrow$ pairs and $M$ is the area where the symmetric mixed ($\uparrow\downarrow+\uparrow\uparrow+\downarrow\downarrow$) phase realizes.}\label{phase_diagram_fig}
\end{figure}

Let us describe all phases more quantitatively. A pure singlet state is characterized by non-zero $\Delta$ (and zero $\eta_{\sigma}$) and in the mean field approximation undergoes the BCS-BEC crossover (sign of the chemical potential changes from positive to the negative one)  at $a_sp_F=\pi/4$ (here $\hbar p_F$ is the Fermi momentum of spin-up or spin-down particles). When the interspecies interaction is turned off $g_s=0$, the triplet condensates of $\uparrow\uparrow$ and $\downarrow\downarrow$ pairs
live at any magnitude and sign of $a_p$. Here the BCS-BEC crossover occurs at $a_pp_F=\pi/4$. An explicit analytic formula for the chemical potential (in units of Fermi energy $\mu_F$)
\begin{eqnarray}\label{mu_sym}
\frac{\mu}{\mu_F}=\frac{1}{a_sp_F}\left[\frac{4}{\pi}-\frac{1}{a_pp_F}\right],
\end{eqnarray}
can be obtained in the symmetric phase, where both pairing mechanisms realize in the system with the triplet condensates being in phase $\eta_{\downarrow}=\eta_{\uparrow}$.
The most complicated, from the point of view of numerical calculations, is the antisymmetric phase (as in the symmetric phase, it is a coexistence of three superfluid components but with order parameters being out of phase $\eta_{\downarrow}=-\eta_{\uparrow}$). The region of its emergence was identified by simultaneous solution of coupled equations (\ref{Omega_min}) and is presented in Fig.~\ref{anti-symm_phase_fig}.
However, the chemical potential of antisymmetric phase in this region is not minimal making this phase of the system at least metastable. The full MF phase diagram of two-component fermions with $s$-wave and $p$-wave interactions is presented in Fig.~\ref{phase_diagram_fig}.
Importantly, there is a room for symmetric phase (regions denoted by $M$ in Fig.~\ref{phase_diagram_fig}), which manifests three-component superfluid. The phase diagram is characterized by two triple points. A location, $(a_sp_F, a_pp_F)=(\pi/4,\pi/4)$, of the first one can be guessed from the above discussion, since exactly at this point chemical potentials of three phases equal zero identically. The second triple point, $(a_sp_F, a_pp_F)=(1.72,2.39)$, is located from the BEC side of the $p$-wave BCS-BEC crossover. It is worth noting that the symmetric mixed phase always survives from the BCS side of the triplet BCS-BEC crossover at weak $s$-wave attraction, and lives in a small `triangle' from the BEC side.

\section{Summary}
In conclusion, we have studied properties of spin-$1\over 2$ Fermi system with equal population of particles in each spin state and a contact two-body interaction described by the $s$-wave and $p$-wave pseudopotentials between fermions in a different and same spin state, respectively. After brief discussion of the two-body problem in dimensions $D<2$ (where all interactions are universal), we have explored, utilizing the mean-field approximation, the superfluid properties of the one-dimensional system in the many-body limit. In particular, we have identified two mixed phase of two-component fermions, where the singlet and triplet Cooper pairs coexist. Our numerical calculations suggest that only one of them -- with equal order parameters of the triplet $\uparrow\uparrow$ and $\downarrow\downarrow$ pairs -- is thermodynamically stable. The detailed phase diagram of the system, which is the main result of this study, is presented in Fig.~\ref{phase_diagram_fig}. It contains two triple points (the highest possible that allowed by the Gibbs phase rule), and a number of the first order quantum phase transition lines. Two interesting questions to be answered in future are the following: an impact of the Gaussian fluctuation of order parameters, and a role of the $p$-wave interaction between particles in different spin states in the phase diagram formation.

\section*{Acknowledgements}
This work was partly supported by Project No.~0122U001514 from the Ministry of Education and Science of Ukraine.

\end{document}